\documentclass[11pt,floatfix,noshowpacs]{revtex4}
\usepackage[dvips]{graphicx,color}

\usepackage[caption=false]{subfig}

\usepackage[latin1]{inputenc}
\usepackage{amsmath,amssymb,amsfonts,amsthm,latexsym}
\usepackage[english]{babel}
\usepackage{float}
\usepackage{subfig}
\usepackage{slashed}

\def\Tr{\mbox{Tr}\,}
\def\openone{\leavevmode\hbox{\small1\kern-3.8pt\normalsize1}}%

\def\mf{{\mbox{\tiny MF}}}
\def\qq{\langle\bar qq\rangle}

\graphicspath{eps/}

\begin{document}

\author{J.P. Carlomagno$^{a,b}$, D. G\'omez Dumm$^{a,b}$ and N.N.\ Scoccola$^{b,c,d}$}

\address{
$^{a}$ IFLP, CONICET $-$ Dpto.\ de F\'isica, Universidad
Nacional de La Plata, C.C. 67, 1900 La Plata, Argentina,\\
$^{b}$ CONICET, Rivadavia 1917, 1033 Buenos Aires, Argentina \\
$^{c}$ Physics Department, Comisi\'on Nacional de Energ\'ia
At\'omica,\\
Av.Libertador 8250, 1429 Buenos Aires, Argentina \\
$^{d}$ Universidad Favaloro, Sol\'is 453, 1078 Buenos Aires, Argentina}

\title{\sc\Large{Inhomogeneous phases in nonlocal chiral quark models}}

\begin{abstract}
The presence of inhomogeneous phases in the QCD phase diagram is
analyzed within chiral quark models that include nonlocal
interactions. We work at the mean field level, assuming that the
spatial dependence of scalar and pseudo-scalar condensates is
given by a dual chiral density wave. Phase diagrams for Gaussian
nonlocal form factors are studied in detail and compared with
those obtained within the Nambu$-$Jona-Lasinio model and
quark-meson approaches.

\end{abstract}

\maketitle


\section{Introduction}

Due to the well-known sign problem, present lattice QCD analyses
are still not able to provide fully faithful predictions for QCD
thermodynamics at low temperatures and relatively high chemical
potentials, including the region where the critical point is
expected to appear. Thus, our knowledge of the strongly
interacting matter phase diagram largely relies on the study of
effective models, which offer the possibility to get predictions
of the transition features at regions that are not accessible
through lattice techniques. In this context, in the last years
some works have considered that the chiral symmetry restoration at
low temperatures could be driven by the formation of nonuniform
phases~\cite{Buballa:2014tba}. One particularly interesting result
suggests that the expected critical endpoint of the first order
chiral phase transition might be replaced by a so-called Lifshitz
point (LP), where two homogeneous phases and one inhomogeneous
phase meet~\cite{Nickel:2009ke}. This result has been obtained in
the chiral limit, where the end point becomes a tricritical point
(TCP), in the framework of the Nambu$-$Jona-Lasinio model
(NJL)~\cite{njl}. As it is well known, in this model quark fields
interact through a local chirally invariant four-fermion coupling.
More recently, this issue has also been addressed in the context
of a quark-meson (QM) model with vacuum
fluctuations~\cite{Carignano:2014jla}, where it is found that the
LP might coincide or not with the TCP depending on the model
parametrization.

In a previous work~\cite{Carlomagno:2014hoa} we have analyzed the
relation between the positions of the TCP and LP in the framework
of nonlocal chiral quark models using a generalized
Ginzburg-Landau approach. It should be mentioned that nonlocal
models can be viewed as extensions of the NJL model that intend to
represent a step towards a more realistic modelling of QCD. In
fact, nonlocality arises naturally in the context of successful
approaches to low-energy quark
dynamics~\cite{Schafer:1996wv,RW94}, and it has been
shown~\cite{Noguera:2008} that nonlocal models can lead to a
momentum dependence in the quark propagator that is consistent
with lattice QCD
results~\cite{bowman,Parappilly:2005ei,Furui:2006ks}. Another
advantage of these models is that the effective interaction is
finite to all orders in the loop expansion, and therefore there is
not need to introduce extra cutoffs~\cite{Rip97}. Moreover, in
this framework it is possible to obtain an adequate description of
the properties of strongly interacting particles at both zero and
finite
temperature/density~\cite{Noguera:2008,Bowler:1994ir,Schmidt:1994di,Golli:1998rf,
GomezDumm:2001fz,Scarpettini:2003fj,GomezDumm:2004sr,GomezDumm:2006vz,
Hell:2008cc,Contrera:2009hk,Hell:2009by,Contrera:2010kz,Dumm:2010hh,
Pagura:2011rt,Carlomagno:2013ona}. The results of
Ref.~\cite{Carlomagno:2014hoa} indicate that for all
phenomenologically acceptable parametrizations considered the TCP
is located at a higher temperature and a lower chemical potential
in comparison with the LP. Consequently, these models seem to
favor a scenario in which the onset of the first order transition
between homogeneous phases is not covered by an inhomogeneous,
energetically favored phase. The aim of the present work is to
further investigate the consequences of the possible existence of
inhomogeneous condensates on the thermodynamics of nonlocal models
by explicitly constructing the associated phase diagrams in the
mean field approximation. In principle, a full analysis would
require to consider general spatial dependent condensates, looking
for the configurations that minimize the mean field thermodynamic
potential at each value of the temperature and chemical potential.
Since for an arbitrary 3-dimensional configuration this turns out
to be a very difficult task, even in the case of local models it
is customary to consider one-dimensional modulations, expecting
that the qualitative features of the inhomogeneous phases will not
be significantly affected by the specific form of the spatial
dependence carried by the condensates~\cite{Buballa:2014tba}. Due
to the additional difficulties introduced by the presence of
nonlocal quark-quark interactions, here we will consider a simple
one-dimensional configuration, namely the so-called dual chiral
density wave (DCDW)~\cite{Nakano:2004cd}, for which the spatial
dependence of the quark condensates is given by
\begin{equation}
\langle \bar q(\vec x) q(\vec x)\rangle \ \alpha \ \cos(\vec Q\cdot\vec x)\ ,
\qquad \qquad
\langle \bar q(\vec x) i \gamma_5 q(\vec x)\rangle \ \alpha \ \sin (\vec Q\cdot\vec x)\ ,
\label{uno}
\end{equation}
for both  $q = u$ and $d$ quark flavors. Regarding the nonlocal
interactions, we will consider the case of covariant and
instantaneous nonlocal form factors with a Gaussian momentum
dependence.

The article is organized as follows. In Sect.~II we present the
general theoretical framework and propose an ansatz for the
bosonic mean fields that leads to the required spatial dependence
of chiral condensates. The model parametrization is also briefly
introduced. Then in Sect.~III we show the phase diagrams for
various parametrizations and discuss the features of the
corresponding phase transitions. Finally, in Sect.~IV we state our
conclusions.


\section{Theoretical framework}

\subsection{Inhomogeneous condensates in nonlocal chiral quark models}

Let us consider a two-flavor model that includes a four-point
coupling between nonlocal quark-antiquark currents. The
corresponding effective action in Euclidean space is given
by~\cite{GomezDumm:2006vz}
\begin{equation}
S_{E}= \int d^{4}x\
\left[ \bar{\psi}(x) \ (-i\, \rlap/\partial\, + \, m_c )\ \psi(x)
- \frac{G}{2} \ j_{a}(x) \ j_{a}(x)  \right] \ ,
\label{action}
\end{equation}
where $\psi$ is the fermion doublet $\psi\equiv(u,d)^T$ and $m_c$
stands for the current quark mass in the isospin limit. The
nonlocal currents $j_{a}(x)$ are given by
\begin{align}
j_{a}(x)  &  =\int d^{4}z\ {\cal G}(z)\
\bar{\psi}\left(x+\frac{z}{2}\right) \ \Gamma_{a}\ \psi\left(
x-\frac{z}{2}\right) , \label{currents}
\end{align}
where we have defined $\Gamma_{a}=(\Gamma_0,\vec\Gamma)=
(\leavevmode\hbox{\small1\kern-3.8pt\normalsize1},
i\gamma_{5}\vec{\tau})$, while ${\cal G}(z)$ is a form factor that
characterizes the effective interaction.

The model can be bosonized through the introduction of bosonic
fields $\Phi_a(x)$ associated to the quark bilinears in
Eq.~(\ref{currents})~\cite{Rip97}. A standard procedure leads to
the Euclidean action
\begin{equation}
S_E \ = \
\int d^4x\; d^4x' \ \bar \psi(x')\; D^{-1}(x',x)\;\psi(x) +
\frac{1}{2G} \int d^4x\; \Phi_a(x)\, \Phi_a(x)\ ,
\label{action}
\end{equation}
where
\begin{equation}
D^{-1}(x',x) \ = \ \delta^{(4)}(x'-x)(-i\rlap/\partial_x\, + \,
m_c) + {\cal G}(x'-x)\, \Gamma_a\; \Phi_a\left(\frac{x+x'}{2}\right) \ .
\label{operx}
\end{equation}
We will work within the mean field approximation (MFA), in which
the bosonic fields are expanded around a real classical
configuration $\bar \Phi_a(x)$. After integrating the fermion
degrees of freedom one obtains the bosonized action
\begin{equation}
S^{\rm (bos)}_\mf\  = \ \int d^4 x \int
d^4 x' \left[ \Tr \log D_\mf^{-1}(x',x) + \frac{1}{2G}\;\bar\Phi_a(x')
\bar\Phi_a(x)\, \delta^{(4)}(x'-x)\right] \ ,
\label{bosonic}
\end{equation}
where the trace acts on Dirac, flavor and color spaces.

Let us consider a system in equilibrium at finite temperature $T$
and chemical potential $\mu$, where inhomogeneous phases could be
favored.  At the mean field level the grand canonical
thermodynamic potential per unit volume is given by
\begin{equation}
\omega_\mf \ = \ -\,\frac{T}{V}\,\log {\cal Z}_\mf\ ,
\label{omegaz}
\end{equation}
where ${\cal Z}_\mf$ is the mean field partition function that
arises from the effective action in Eq.~(\ref{bosonic}). If the
ground state is in general not homogeneous, the quark condensate
at a given position $\vec x$ can be calculated by introducing an
auxiliary static field $\varphi(\vec x)$. One has
\begin{equation}
\langle \bar \psi(\vec x)\,\psi(\vec x) \rangle \ = \ -\,
\frac{\delta \log {\cal Z}[\varphi]}{\delta \varphi(\vec
x)}\Bigg|_{\varphi = 0} \ ,
\label{condenx}
\end{equation}
where ${\cal Z}[\varphi]$ is obtained from ${\cal Z}_\mf$ by
changing $D^{-1}_\mf(x',x) \to D^{-1}_\mf(x',x)+\delta^{(4)}(x'-
x)\,\varphi(\vec x)$ in the inverse propagator given in
Eq.~(\ref{operx}), taken at mean field. Moreover, since now parity
is not necessarily an exact symmetry of the vacuum, one can get in
general a nonzero value for the condensate $\langle \bar \psi(\vec
x)\, \Gamma_3 \psi(\vec x)\rangle = \langle \bar \psi(\vec x)\, i
\gamma_5\tau_3 \psi(\vec x)\rangle$. The latter can be obtained
from the partition function by adding a term
$\delta^{(4)}(x'-x)i\gamma_5\tau_3\varphi(\vec x)$ to the inverse
propagator in Eq.~(\ref{operx}) at mean field.

The thermodynamics can be worked out using the Matsubara
formalism. Thus, it is convenient to consider the inhomogeneous
mean field propagator in momentum space. One has
\begin{equation}
D_\mf(p',p) = \left[(-\slashed p + m_c)\;(2\pi)^4\,
\delta^{(4)}(p'-p)\; + \; g\left(\frac{p+p^\prime}{2}\right)\
\Gamma_a \,\bar\Phi_a(p'-p)\right]^{-1} \ ,
\label{oper}
\end{equation}
where $g(p)$ and $\bar \Phi_a(p)$ are the Fourier transforms of
the form factor ${\cal G}(x)$ and the mean fields $\bar
\Phi_a(x)$, respectively. Since energy is conserved for static
mean field configurations, it is also useful to define a reduced
effective propagator $\tilde D(p',p)$ through
\begin{equation}
D_\mf(p',p) \ = \ (2\pi) \delta(p'_4 - p_4) \, \tilde
D(p',p) \ .
\end{equation}
With these definitions the condensates are found to be given by
\begin{eqnarray}
\langle \bar \psi(\vec x)\,\Gamma_a\,\psi(\vec x) \rangle & = & -\,
T\;\sum_{n = -\infty}^{\infty} \int \;\frac{d^3p}{(2\pi)^3}
\;\frac{d^3p'}{(2\pi)^3} \; e^{i(\vec p\,'-\vec p)\cdot\vec x}\;
\Tr [\Gamma_a\tilde D(p',p)] \ , \quad a = 0,3\ ,\label{trazas}
\end{eqnarray}
where the traces are taken over Dirac, flavor and color spaces. Here the
fourth component of $p'$ and $p$ in $\tilde D(p',p)$ is given by $p'_4 = p_4
= \omega_n - i \mu$, where $\mu$ is the chemical potential and $\omega_n =
(2n+1)\pi T$ are the fermionic Matsubara frequencies.

\subsection{Dual chiral density wave}

We address here the relatively simple situation in which the vacuum is
modulated by a dual chiral density wave. In this configuration the chiral
condensate rotates along the chiral circle, carrying a constant
three-momentum $\vec Q$ [see Eq.~(\ref{uno})]. For simplicity, in the
following we will consider the case of vanishing current quark masses, $m_c
= 0$. In this limit, the desired behavior of the chiral condensates can be
obtained by considering the following ansatz for the mean field
configuration $\bar\Phi_a(p'-p\,;\vec Q)$~\cite{Muller:2013tya}:
\begin{equation} \Gamma_a\,\bar\Phi_a(p'-p\,;\vec Q) \ =
(2\pi)^4\,\delta(p'_4-p_4)\ \phi\;\;\sum_{s=\pm} \;
\frac{1+s\,\gamma_5\tau_3}{2}\;\delta^{(3)}(\vec p\,'-\vec p+s\,
\vec Q)\ .
\end{equation}
From this ansatz it is evident that the effective propagator will
be block diagonal in flavor space, thus it can be written as a
direct sum of $D_u$ and $D_d$ propagators. By calculating the
inverse in Eq.~(\ref{oper}) we get for the $u$ quark
\begin{equation}
\tilde D_u(p',p\,;\vec Q) \ = \
\left( \begin{array}{cc} B_+(\frac{p'+p}{2}\,;\vec Q)
\;\delta^{(3)}(\vec p\,'-\vec p-\vec Q) &
A_-(p\,;\vec Q)\;\delta^{(3)}(\vec p\,'-\vec p)
\\ A_+(p\,;\vec Q)\;\delta^{(3)}(\vec p\,'-\vec p) &
B_-(\frac{p'+p}{2}\,;\vec Q)\;\delta^{(3)}(\vec p\,'-\vec p+\vec Q)
\end{array} \right) \ ,
\end{equation}
where $A_\pm(p\,;\vec Q)$ and $B_\pm(\frac{p'+p}{2}\,;\vec Q)$ are $2\times
2$ matrices in Dirac space. These are given by
\begin{eqnarray}
A_\pm(p\,;\vec Q) & = & \frac{1}{\Delta(p\,;\vec Q)}\left\{[p_4^2
+ (\vec p \pm \vec Q)^2 + \phi^2 g(p)^2] (i p_4 \openone\pm \vec p
\cdot \vec \tau) + \phi^2 g(p)^2 \vec Q\cdot \vec\tau\right\}
 \nonumber  \\
B_\pm(t\,;\vec Q) & = & \frac{\phi\, g(t)}{\Delta(t\,;\vec Q)}\left\{ [t^2 -
\vec Q\,^2/4 + \phi^2 g(t)^2] \openone - i (\vec t \times \vec Q \pm
t_4\,\vec Q) \cdot \vec \tau] \right\} \ ,
\end{eqnarray}
where
\begin{equation}
\Delta(p\,;\vec Q) \ = \ \left[p^2 - \vec Q\,^2/4 + \phi^2 g(p)^2\right]^2
+ p^2 \vec Q\,^2 - (\vec p\cdot \vec Q)^2\ .
\end{equation}
The mean field propagator for the $d$ quark is obtained from the
previous expressions by
\begin{equation}
\tilde D_d(p',p\,;\vec Q) \ = \ \tilde D_u(p',p\,;-\vec Q) \ .
\end{equation}
In this way, from Eq.~(\ref{trazas}) we obtain
\begin{eqnarray}
\langle \bar u(\vec x) u(\vec x)\rangle & = &
\langle \bar d(\vec x) d(\vec x)\rangle \ = \ F(Q^2) \cos(\vec Q\cdot \vec x)\ ,
\nonumber \\
\langle \bar u(\vec x)i\gamma_5 u(\vec x)\rangle & = & - \langle
\bar d(\vec x)i\gamma_5 d(\vec x)\rangle \ = \ F(Q^2) \sin (\vec Q\cdot \vec x)
\ ,
\label{condenq}
\end{eqnarray}
where
\begin{equation}
F(Q^2) \ = \ -\,4\, N_c \,T\;\sum_{n = -\infty}^{\infty}
\int \;\frac{d^3p}{(2\pi)^3}\; \frac{\phi \, g(p)
\, [p^2 - Q^2/4 + \phi^2 g(p)^2]}{\Delta(p\,;\vec Q)} \ ,
\label{fq2}
\end{equation}
with $p_4 = \omega_n - i\mu$.

If the ground state is assumed to be homogeneous, the mean fields
$\bar\Phi_a(x)$ are uniform. Then, from parity invariance one has
$\bar\Phi_a(p) = (2\pi)^4\delta^{(4)}(p)\, \delta_{a0}\ \phi$, and
the operator in Eq.~(\ref{oper}) can be trivially inverted. The
corresponding expressions for the condensates are obtained in this
case by setting $\vec Q = 0$ in Eqs.~(\ref{condenq}) and
(\ref{fq2}), namely~\cite{GomezDumm:2001fz,GomezDumm:2006vz}
\begin{eqnarray}
\langle \bar u u\rangle & = & \langle \bar d d\rangle \
= \ -\,4\, N_c \,T\;\sum_{n = -\infty}^{\infty}
\int \;\frac{d^3p}{(2\pi)^3}\; \frac{g(p)\,\phi}
{p^2 + g(p)^2\,\phi^2 }\ , \nonumber \\
\langle \bar u i \gamma_5\tau_3 u\rangle & = & \langle \bar d
i \gamma_5\tau_3 d\rangle \ = \ 0 \ .
\end{eqnarray}

Let us now evaluate the thermodynamic potential in Eq.~(\ref{omegaz}) for
the case of the dual chiral density wave. From the mean field partition
function ${\cal Z}_\mf (T,\mu)$, using the Matsubara formalism we obtain the
grand canonical thermodynamic potential per unit volume
\begin{equation}
\omega_\mf(T,\mu) \ = \
\ -\,2\, N_c\,T\sum_{n = -\infty}^{\infty}
\int \frac{d^3p}{(2\pi)^3}\;\log\Delta(p\,;\vec Q)
\ + \frac{\phi^2}{2G}\ ,
\end{equation}
where once again the fourth component of $p$ in $\Delta(p\,;\vec
Q)$ is $p_4 = \omega_n - i\mu$. Here the integral over $p$ is
divergent for large momenta. A standard way of regularizing the
thermodynamic potential is by subtracting the free contribution
$\omega_{\rm free} = \omega_\mf(\phi = 0)$ and adding it in a
regularized form. In this way one ends up with
\begin{equation}
\omega^{\rm reg}_\mf \ = \
-2 N_c \,T\sum_{n = -\infty}^{\infty}
\int \frac{d^3p}{(2\pi)^3}\; \log
\left[ 1 + \phi^2\, g(p)^2\; \frac{\phi^2\, g(p)^2 +
2\,(p^2-Q^2/4)}{(p^2+Q^2/4)^2-(\vec p\cdot \vec Q)^2} \right]
 + \ \frac{\phi^2}{2 G}\ + \ \omega^{\rm reg}_{\rm free}\ ,
\label{om_reg}
\end{equation}
where
\begin{equation}
\omega^{\rm reg}_{\rm free}\ = \ -\ N_c \left[\frac{7\pi^2\,
T^4}{90}\; + \;\frac{T^2\,\mu^2}{3}\; + \; \frac{\mu^4}{6\pi^2}
\right]\ .
\end{equation}
The mean field values $\phi$ and $Q\equiv |\vec Q|$ can be obtained by
looking for the minimum of $\omega_\mf$ through the coupled equations
\begin{equation}
\frac{\partial \omega^{\rm reg}_\mf}{\partial \phi} = 0\ ,\qquad\qquad
\frac{\partial \omega^{\rm reg}_\mf}{\partial Q} = 0 \ .
\label{qys}
\end{equation}
A region in which the absolute minimum is reached for a nonzero
$Q$ will correspond to an inhomogeneous phase. As expected, if
chiral symmetry is not dynamically broken (i.e.~$\phi = 0$) the
regularized thermodynamic potential reduces to the free
contribution $\omega_{\rm free}^{\rm reg}$, which does not depend
on $Q$.


\subsection{Model parameterization}

In the chiral limit the model has only one coupling parameter, namely the
constant $G$. In addition, one has to specify the functional form of the
form factor $g(p)$, which requires the introduction of some momentum scale
$\Lambda$ in order to satisfy Lorentz invariance. For definiteness we will
consider here a Gaussian behavior
\begin{equation}
g(p)= \exp \left(-p^{2}/\Lambda^{2}\right) \ ,
\label{ff1}
\end{equation}
which guarantees a fast ultraviolet convergence of loop integrals.

Given the form factor shape, one can fix the model parameters $G$
and $\Lambda$ so as to reproduce the phenomenological values of
the pion decay constant $f_\pi$ and the chiral quark condensate
$\qq$. In fact, since we are working in the chiral limit, it is
obvious from dimensional analysis that any dimensionless quantity
turns out to be just a function of the dimensionless combination
$\bar G = G \Lambda^2$, while a dimensionful quantity can be
written as a function of $\bar G$ times some power of a
dimensionful parameter, say e.g.\ $f_\pi$. According to the recent
analysis in Ref.~\cite{Aoki:2013ldr}, we will take here
$f_\pi^{\rm ch}=86$~MeV and $\langle\bar q q\rangle^{\rm ch} =
-(270$~MeV$)^3$ (superindices stress that the values correspond to
the chiral limit), thus the ``physical'' value of $\bar G$ will be
that leading to a ratio $(-\langle\bar q q\rangle^{\rm ch}
)^{1/3}/f_\pi^{\rm ch} \simeq 3.14$. In order to check the
parameter dependence of our results we will consider values for
this ratio in the range $2.8$ to $3.5$. For $f_\pi^{\rm
ch}=86$~MeV, this corresponds to a shift of at most $\sim 30$~MeV
around the central value $(-\langle\bar q q\rangle^{\rm ch})^{1/3}
= 270$~MeV.

\begin{figure}[hbt]
\centering
\subfloat{
\includegraphics[scale=0.38]{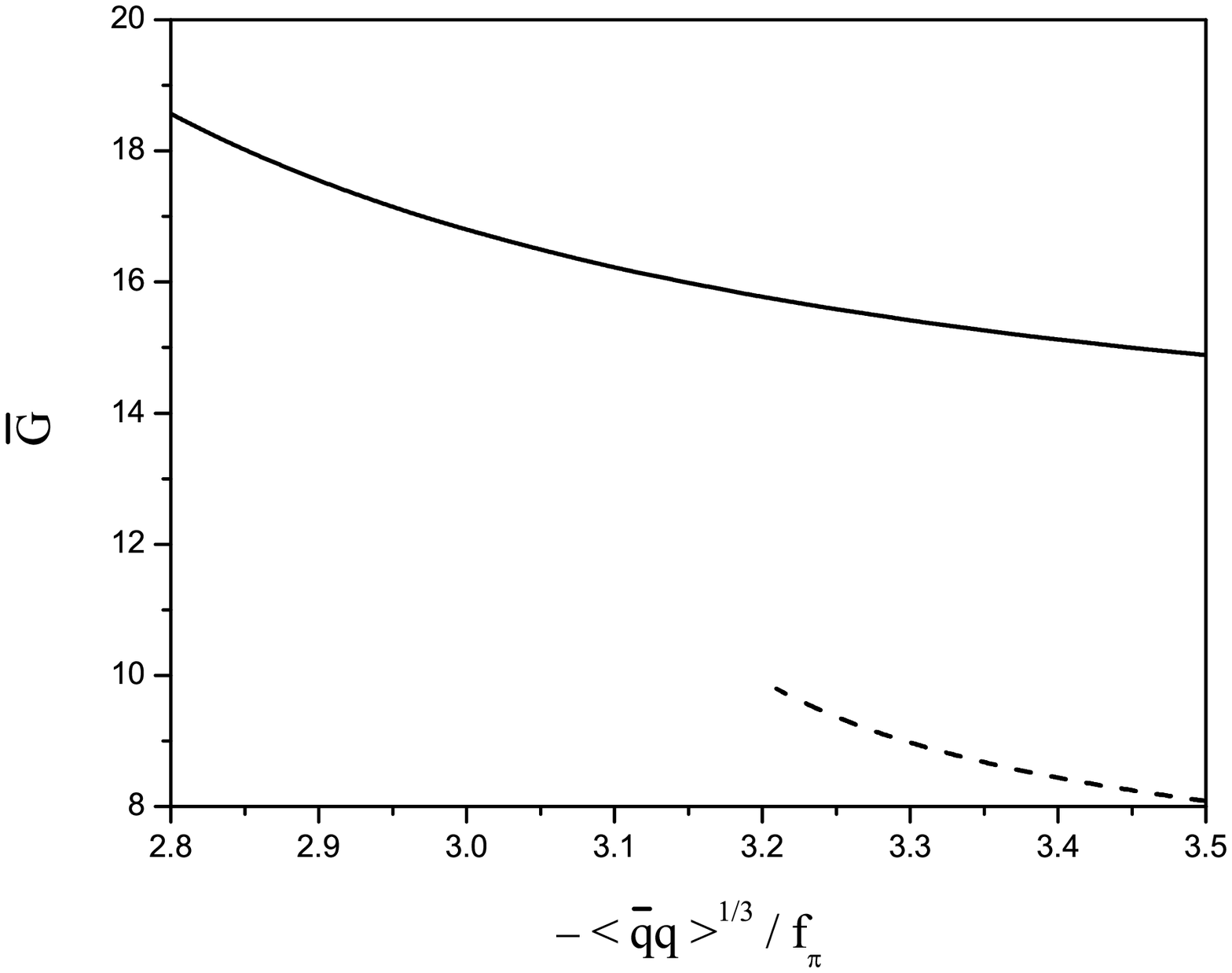}
}
\subfloat{
\includegraphics[scale=0.38]{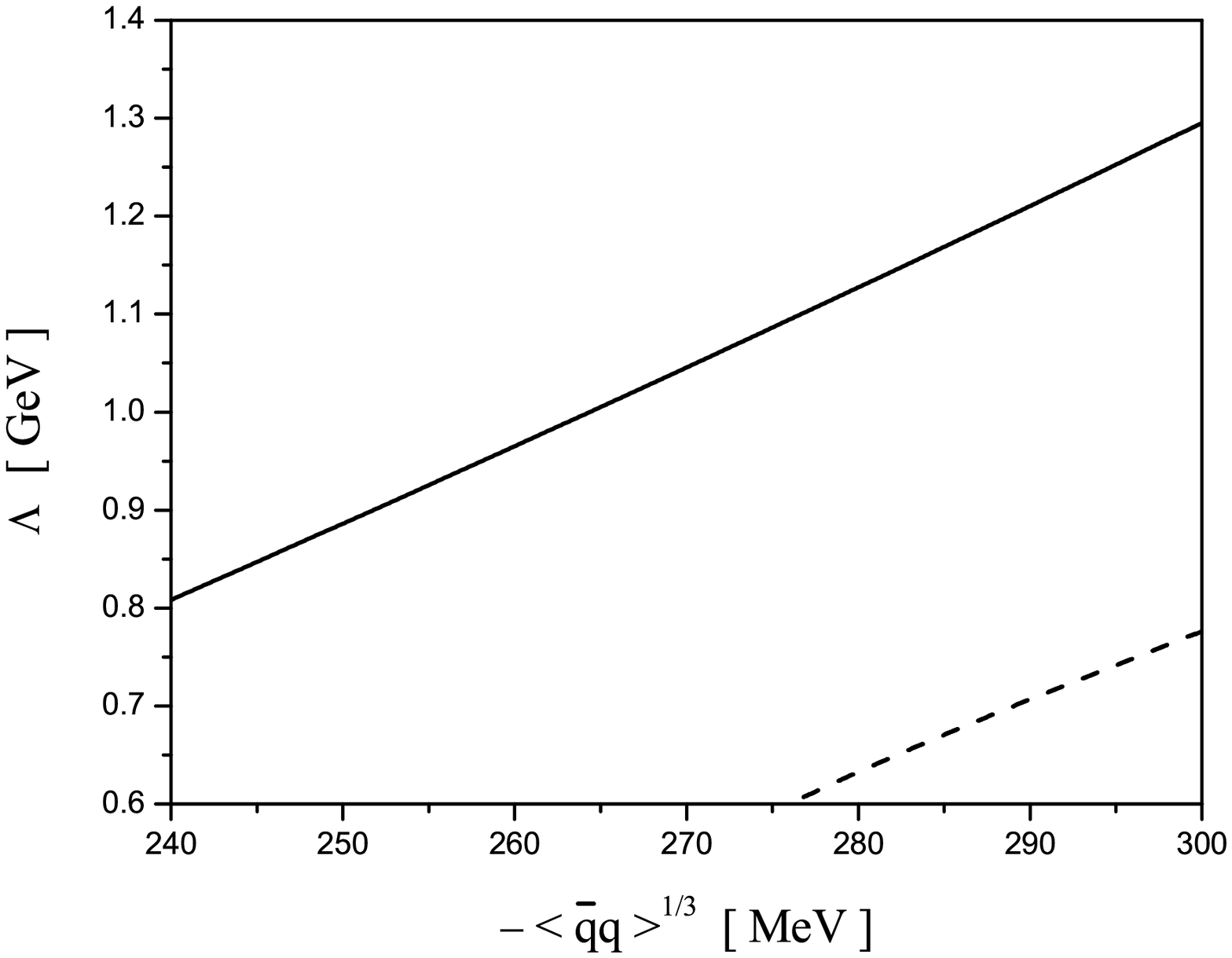}
}
\caption{Left: values of the dimensionless parameter $\bar G$ for
a given ratio $(-\langle \overline{q} q \rangle^{\rm ch})^{1/3} \
/ f_\pi^{\rm ch}$. Right: effective cutoff scale $\Lambda$ for a
given value of the quark condensate, with $f_\pi$ fixed to the
phenomenological value (in the chiral limit) 86~MeV. Solid
(dashed) lines correspond to covariant (instantaneous) Gaussian
form factors.
}
\label{fig:1}
\end{figure}

In Fig.~\ref{fig:1} the solid lines indicate our numerical results
for the parameters $\bar G$ and $\Lambda$ that correspond to the
above range of the ratio $(-\langle \overline{q} q \rangle^{\rm
ch})^{1/3} \ / f_\pi^{\rm ch}$. In the left panel we show the
values of the dimensionless parameter $\bar G$ as a function of
the ratio $(-\langle \overline{q} q \rangle^{\rm ch})^{1/3} \ /
f_\pi^{\rm ch}$, while in the right panel we quote the effective
cutoff scale $\Lambda$ as a function of the quark condensate for
the phenomenologically preferred value $f_\pi = 86$~MeV. The
cutoff values are found to be of order $\sim 1$~GeV, in agreement
with phenomenological expectations.

Finally, it is interesting to consider the case of the so-called
``instantaneous'' form factors, which just depend on the
three-momentum $\vec p$~\cite{Schmidt:1994di}. In this case
Lorentz symmetry is broken, and a spatial cutoff is needed [notice
that, in particular, the usual ``local'' NJL model is obtained by
setting $g(\vec p) = \theta(\Lambda_{\rm NJL}^2-\vec p\,^2)$]. If
we consider once again a Gaussian shape for the form factor,
namely $g(\vec p) = \exp(\vec p\,^2/\Lambda^2)$, and same
phenomenological requirements as in the covariant case, the
corresponding numerical values for $\bar G$ and $\Lambda$ are
those shown by the dashed lines in Fig.~1. As it is discussed in
Refs.~\cite{GomezDumm:2006vz,Grigorian:2006qe}, instantaneous
models with soft cutoff functions lead to relative large values of
the quark condensate, therefore for these models low values of the
cutoff are typically required. We have taken $\Lambda = 600$~MeV
as a lower bound, which implies $(-\langle \overline{q} q
\rangle^{\rm ch})^{1/3} \ / f_\pi^{\rm ch} \geq 3.21$.

\section{Phase Diagrams}

Let us consider a hadronic system at finite temperature $T$ and
chemical potential $\mu$. We will discuss the features of the phase
transitions in the $T - \mu$ plane for the nonlocal chiral quark
models discussed in the preceding section. As stated in the
Introduction, in general one can find phases in which the chiral
symmetry is either broken or restored, and regions in which either
homogeneous or inhomogeneous phases are preferred. In addition, a
general analysis based on the Ginzburg-Landau expansion shows that
nonlocal models allow the presence of both a tricritical point
(TCP) and a Lifshitz point (LP), located in different
positions~\cite{Carlomagno:2014hoa}.

We start by discussing the case of Gaussian covariant nonlocal
form factors. Our numerical results for the corresponding phase
diagrams are displayed in Fig.~2, where we show different
scenarios that may arise if the model parameters lie within the
range discussed in the previous section. We have chosen four
parameter sets, denoted as PI, PII, PIII and PIV, which correspond
to a pion decay constant $f_\pi^{\rm ch} = 86$~MeV and quark
condensate values $(-\qq^{\rm ch})^{1/3} = 240$, 247, 270 and
300~MeV, respectively, at zero $T$ and $\mu$. The different
regions of the phase diagram, as well as the corresponding
transition curves and critical points, are shown in the left
panels of Fig.~2. It is seen that in all cases at low temperatures
and chemical potentials one finds the usual homogeneous, chirally
broken (HCB) phase, while for low $T$ and high $\mu$ the system
lies in an inhomogeneous (IH) phase (in the case of PIV, which
corresponds to $\qq^{\rm ch} = -(300$~MeV$)^3$, the onset of the
IH phase at $T=0$ occurs at a chemical potential of about 630~MeV,
therefore it is not shown in the figure).
\begin{figure}[hbt]
\centering \includegraphics[scale=0.18]{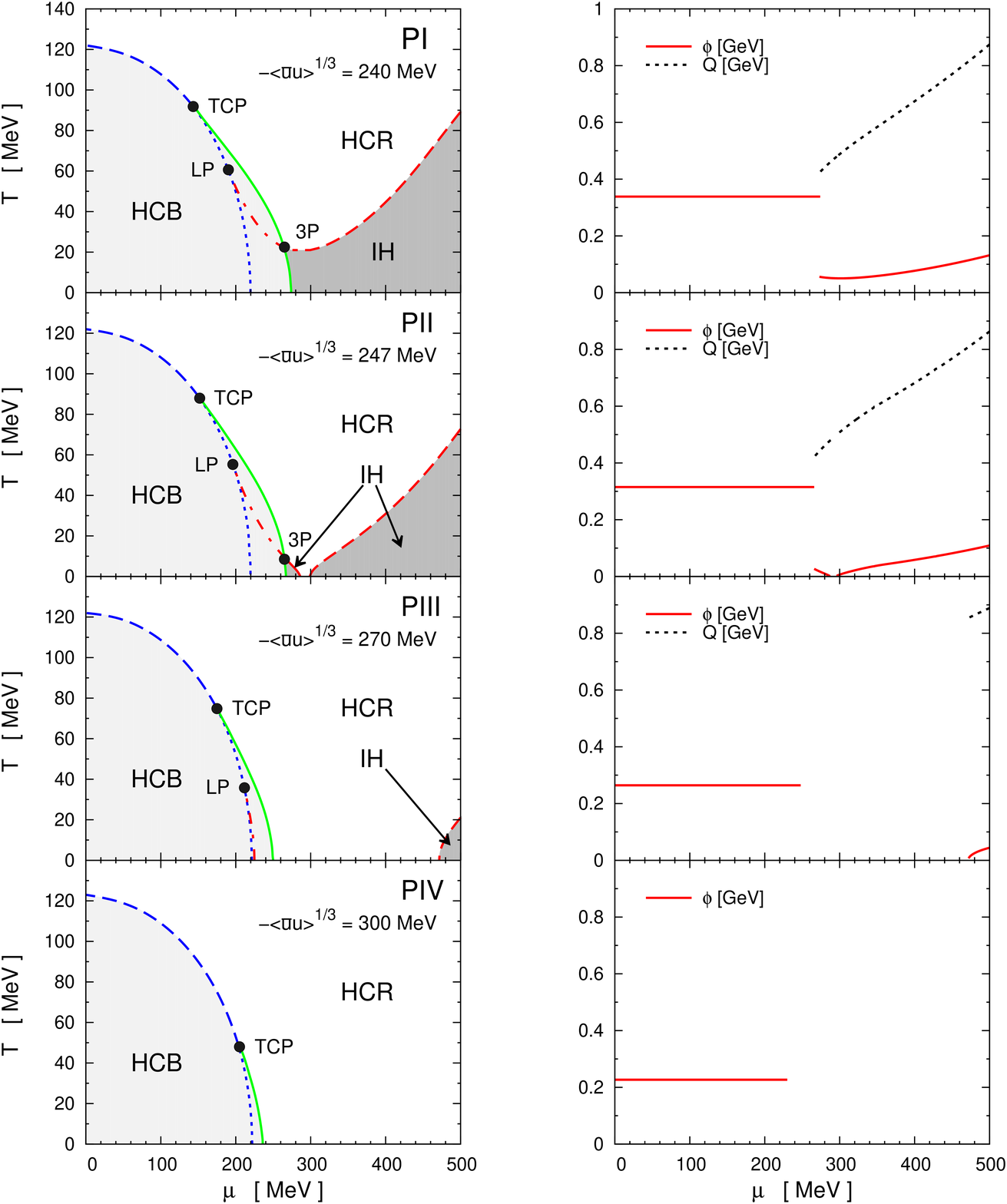}
\caption{Left: $T-\mu$ phase diagrams for different model
parameterizations. Solid (dashed) lines indicate first (second)
order phase transitions. The dotted line is the lower spinodal
corresponding to the homogeneous chiral restoration transition,
while the dashed-dotted line is a boundary of a region in which
there exists a local inhomogeneous minimum of the thermodynamical
potential. TCP, LP and 3P stand for tricritical, Lifshitz and
triple points. Right: values of $\phi$ and $Q$ as functions of the
chemical potential, for $T=0$.} \label{fig:2}
\end{figure}

Let us start by analyzing the case of the parameterization PI (upper left
panel of Fig.~2, corresponding to a relatively low chiral quark condensate
$\qq^{\rm ch} = -(240$~MeV$)^3$, and a relatively high value of the
coupling $\bar G$). The different phases are indicated by the shaded areas,
while solid and dashed lines correspond to first and second order
transitions, respectively. At a temperature of about 100~MeV and low
chemical potentials, it is seen that the system lies in the HCB phase, in
which chiral symmetry is spontaneously broken. As usual, by increasing $\mu$
one finds a second order phase transition to an homogeneous phase in which
chiral symmetry is restored (HCR phase). If the temperature is lowered, the
corresponding second order transition curve ends at a tricritical point,
beyond which it becomes a first order transition line. Now, by following
this line, at a temperature $T_{3P} \simeq 20$~MeV one arrives at a triple
point. For $T < T_{3P}$, at a given critical chemical potential $\mu_c(T)$
the system undergoes a first order transition from the HCB phase into an IH
phase, in which chiral symmetry is found to be only approximately restored.
On the other hand, if one starts with a system in the IH phase and increases
the temperature at constant chemical potential, at some critical value of
$T$ one arrives at a second order phase transition into the HCR phase. As it
is shown in the figure, the corresponding second order transition line
continues beyond the triple point with a dashed-dotted line inside the HCB
area. The latter represents a boundary of a region in which the
thermodynamic potential has a local minimum that corresponds to an
(unstable) IH phase. Finally, in the phase diagram we also show with a
dotted line the lower spinodal corresponding to the homogeneous chiral
restoration transition.

The previously described first order transition from the HCB to
the IH phase is illustrated in Fig.~3: left and right panels show
contour plots of the mean field thermodynamic potential
$\omega_\mf^{\rm reg}(\phi,Q)$ at zero temperature for $\mu = 260$
and $\mu = 280$~MeV, respectively, which correspond to both sides
of the transition point $\mu_c(0) = 274$~MeV. The plots clearly
show the transition from an absolute minimum at $\phi \simeq
340$~MeV, $Q = 0$, to another one in which $\phi$ reduces to about
50~MeV, while the chiral condensates get spatial dependencies as
those given by Eq.~(\ref{condenq}), with $Q \simeq 450$~MeV. These
features are also shown in the upper right panel of Fig.~2, where
we quote the curves for $\phi$ and $Q$ at $T=0$ as functions of
the chemical potential. Notice that on the HCB side (left panel of
Fig.~3) there also exists a local minimum at $(\phi,Q)\sim (50$
MeV, 400 MeV).
\begin{figure}[htb]
\centering
\subfloat{
\includegraphics[scale=0.4]{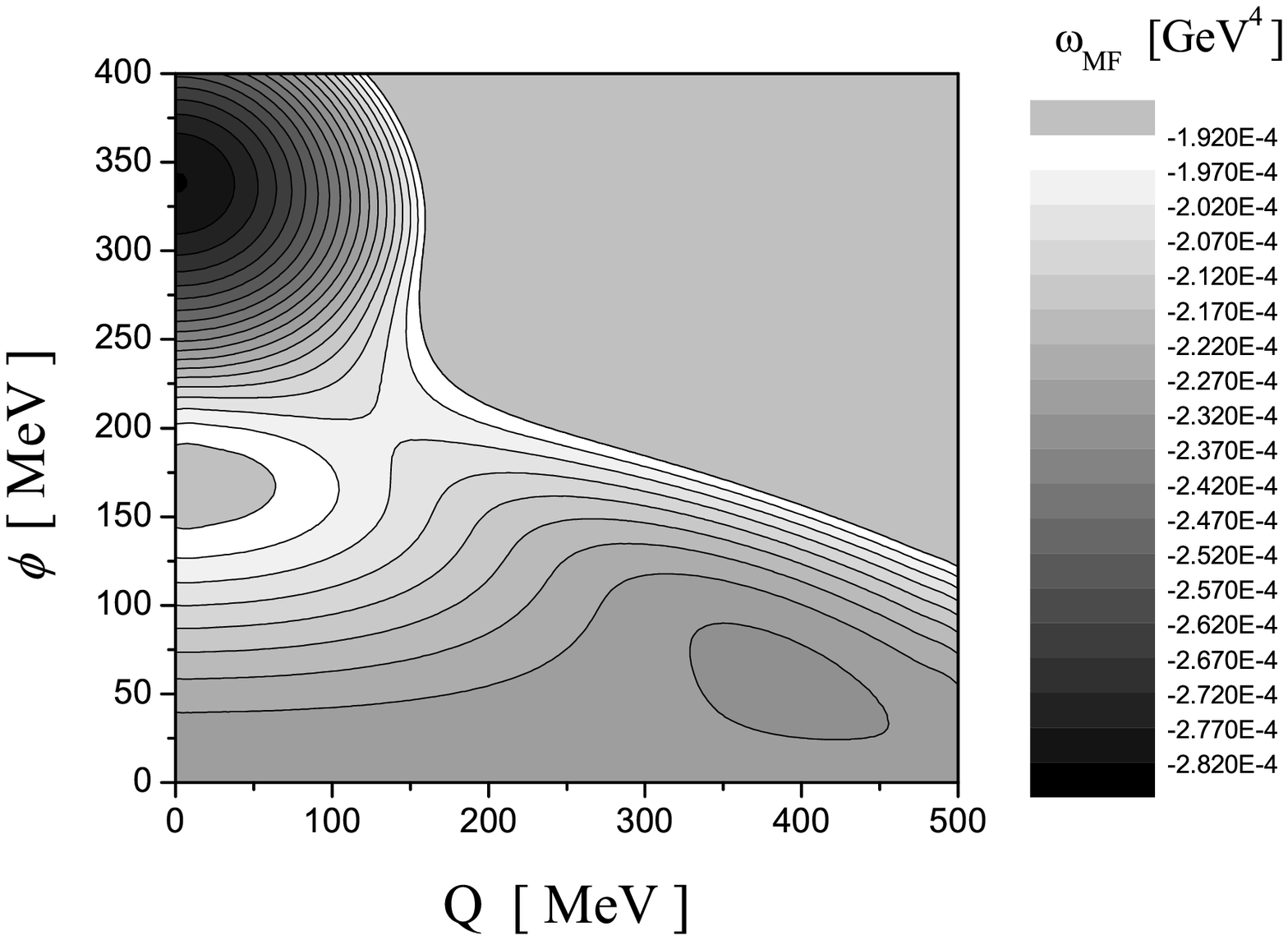}
}
\subfloat{
\includegraphics[scale=0.4]{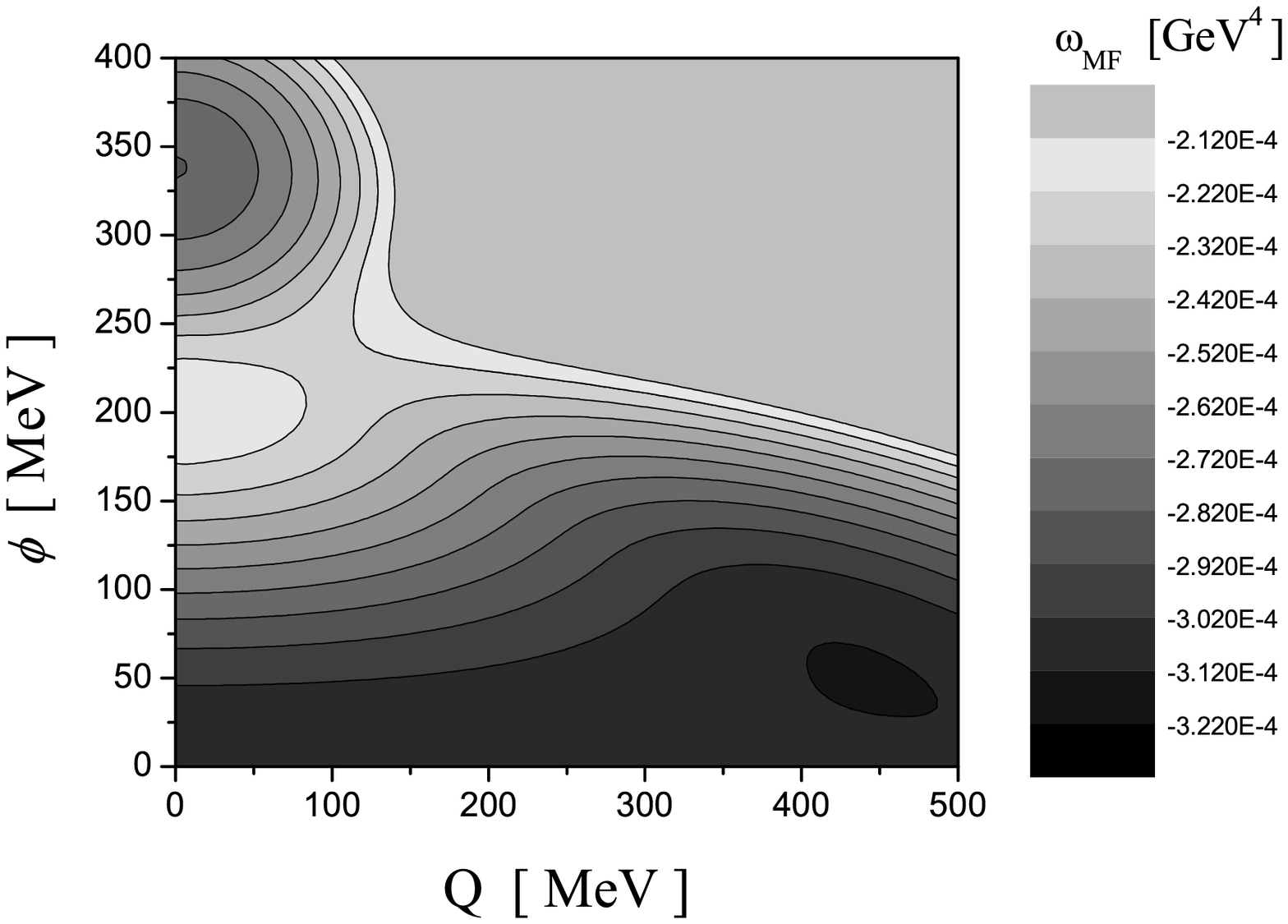}
} \caption{Contour plots of the thermodynamic potential
$\omega^{\rm reg}_\mf$ for a nonlocal chiral quark model at zero
temperature and finite chemical potential, close to the first
order transition between HCB and IH phases. The plots correspond
to parameterization PI, for chemical potentials $\mu = 260$~MeV
(left) and $\mu = 280$~MeV (right).}
\end{figure}

Below the previously discussed phase diagram we show in Fig.~2 the
case of parameterization PII, in which the quark condensate at
zero $T$ and $\mu$ is slightly larger, namely $\qq^{\rm
ch} = -(247$~MeV$)^3$. It can be seen that in this case the IH
phase region gets reduced and splits in two: a small ``island'' of
IH phase becomes isolated from the large IH phase ``continent''
found at high chemical potentials (see shaded regions in the
figure). Then, for PIII and PIV it is seen that the island
disappears, and the onset of the continent is pushed up to larger
values of the chemical potential. The discontinuity of $Q$ at this
transition for $T=0$ becomes increased, as it is shown in the
right panels of Fig.~2. As stated above, for PIV (lower panels of
Fig.~2) it is found that the inhomogeneous phase occurs for values
of $\mu$ out of the region of the phase diagram displayed in our
graphs. For comparison, in Table 1 we quote the values of the
effective cutoffs, the values of $\phi$ and the chiral condensate
$(-\qq^{\rm ch})^{1/3}$ for zero $T$ and $\mu$, and the critical
chemical potentials at $T=0$, for parameterizations PI to PIV. We
denote by $\mu_c'(0)$ the onset of the IH phase continent, while
$\mu_c''(0)$ stands for the chemical potential at the transition
between the IH phase island and the HCR phase region. All values
are given in MeV.
\begin{table}
\begin{tabular}{c c c c c c c }
\hline
     & $(-\qq^{\rm ch})^{1/3}$ & \ \ \ $\Lambda$ \ \ \ & \ \ $\phi$ \ \ &
    \ $\mu_c(0)$ \ & \ $\mu_c''(0)$ \ & \ $\mu_c'(0)$ \ \\
\hline
\hline
PI   & 240 & 808 & 338 & 274 & - & - \\
PII  & 247 & 863 & 315 & 266 & 288 & 295 \\
PIII & 270 & 1045 & 264 & 249 & - & 470 \\
PIV  & 300 & 1295 & 227 & 236 & - & 629 \\
\hline
\end{tabular}
\label{table:} \caption{Effective cutoff $\Lambda$, chiral condensate and
mean field $\phi$ at zero $T$ and $\mu$, and zero-temperature critical
chemical potentials for parametrizations PI to PIV. All values are given in
MeV.}
\end{table}

It is worth pointing out that, for these models, the would-be
Lifshitz point (i.e.~the point where the HCB and HCR phases would
meet the IH one along the second order phase transition lines) is
hidden inside the HCB phase region: around the would-be LP, the
HCB phase turns out to be energetically preferred in all cases
considered. Instead, as it is indicated in the upper left panels
of Fig.~2, a triple point can be found in the case of PI and PII.
It is also worth mentioning that the second order phase transition
curves, as well as both the TCP and would-be LP, can be calculated
for these models through a quite precise semianalytical
approach~\cite{GomezDumm:2004sr,Carlomagno:2014hoa}.

The characteristics of the phase diagrams can be compared with those
obtained within the NJL and the quark-meson model, which have been recently
analyzed in this context. As stated in the Introduction, in the NJL the TCP
and LP are coincident, whereas in the QM this can be the case or not,
depending on the parameterization. In some cases it is
shown~\cite{Carignano:2014jla} that the HCB-HCR second order phase
transition ends at a Lifshitz point, while the TCP appears to be hidden into
the IH region (i.e., the opposite situation to that found in our models). It
is interesting to notice that, according to the analyses in
Refs.~\cite{Carignano:2011gr,Carignano:2014jla,Buballa:2014tba}, for both
the NJL and QM models some parameterizations lead to phase diagrams that
show IH ``continents'' that extend to arbitrarily high chemical potentials.
In fact, it is a matter of discussion whether the presence of these
continents arises just as a regularization artifact. We stress that in
nonlocal models the ultraviolet convergence of loop integrals follows from
the behavior of form factors, which effectively embrace the underlying QCD
interactions (indeed, the form factors can be fitted from lattice QCD
calculations for the effective quark
propagators~\cite{Noguera:2008,Carlomagno:2013ona}). The fact that
various quark models including different regularization procedures lead to
similar qualitative features of the phase diagram seems to indicate that
these features are rather robust. However, it is necessary to mention that
we have not considered the effects of color superconductivity, which are
expected to be important at intermediate and large chemical potentials and
could have a significant impact in the phase diagram.

Finally, we address the case of instantaneous form factors mentioned in the
previous section. For these parametrizations the qualitative features of the
phase diagrams are found to be basically the same as those discussed above.
The main difference is that similar diagrams correspond to models leading to
larger values of the quark condensates: for a model with $(-\qq^{\rm
ch})^{1/3} = 270$~MeV one obtains a phase diagram similar to that of
parametrization PI for the covariant case, while the small ``island'' of
inhomogeneous phase arises when the corresponding condensate is $(-\qq^{\rm
ch})^{1/3} \simeq 285$~MeV. For larger absolute values of the quark
condensates the island disappears and the onset of the inhomogeneous phase
is pushed up to larger values of the chemical potential, just as in the case
of covariant form factors.


\section{Summary and Conclusions}

In this work we have analyzed the possible existence of inhomogeneous phases
in the context of a simple version of nonlocal SU(2) chiral quark models in
the chiral limit. For simplicity, only the one-dimensional modulation
associated to a dual chiral density wave (DCDW) has been considered. In this
framework, different parametrizations of the nonlocality, including both
covariant and instantaneous form factors, have been investigated.

For all studied scenarios it is seen that the sizes of
inhomogeneous phase regions show a rather strong dependence on
model parameters. In all cases we find the existence of a
tricritical point, while, keeping $f_\pi$ fixed, for high values
of the dimensionless coupling $\bar G$ (low absolute values of the
chiral condensate) we find at low temperatures a first order
transition between the homogeneous chirally broken phase and the
inhomogeneous phase. These phases and the homogeneous chirally
restored one meet then at a triple point. On the other hand, for
lower values of $\bar G$ the onset of the inhomogeneous phase is
pushed up to higher chemical potentials and the triple point
disappears. As in previous analyses made in the framework of NJL
and QM models
\cite{Carignano:2011gr,Carignano:2014jla,Buballa:2014tba}, the
inhomogeneous ``continents'' in our phase diagrams extend to
arbitrarily high chemical potentials. Thus, their existence seems
to be a rather robust prediction of this type of quark models. It
should be mentioned, however, that effects of color
superconductivity, which are expected to be important at
intermediate and large chemical potentials, have not been included
in these works. In this sense, it is clear that to clarify this
issue color superconducting interaction channels have to be
incorporated in future calculations. In addition, the role of
vector channels within the framework of the nonlocal models
deserves further investigation as well. Finally, it would be
interesting to explore the possibility of going beyond the DCDW
ansatz used in the present nonlocal scheme by considering more
general one dimensional modulations.

\section*{Acknowledgements}

This work has been partially funded by CONICET (Argentina) under
grants PIP 00682 and PIP 00449, and by ANPCyT (Argentina) under
grant PICT11-03-00113. Financial aid has also been received from
Universidad Nacional de La Plata, Argentina, project \# 11/X718.


\end{document}